\documentclass{vldb}

\usepackage{color}
\usepackage{xcolor}
\usepackage{graphicx}
\usepackage{balance}  % for  \balance command ON LAST PAGE  (only there!)
\usepackage{url}
\usepackage{graphicx}
\usepackage{subfigure}
\usepackage{tabularx}
\usepackage{booktabs}
\usepackage{listings}

\vldbTitle{BlockLite: A Lightweight Emulator for Public Blockchains}
\vldbAuthors{Xinying Wang, Abdullah Al-Mamun, Hui Lin, Feng Yan, Mohammad Sadoghi, and Dongfang Zhao}
\vldbDOI{https://doi.org/10.14778/xxxxxxx.xxxxxxx}
\vldbVolume{12}
\vldbNumber{xxx}
\vldbYear{2019}

\newcommand{\figspace}{\vspace{-3mm}}

\definecolor{codegreen}{rgb}{0,0.6,0}
\definecolor{codegray}{rgb}{0.5,0.5,0.5}
\definecolor{codepurple}{rgb}{0.58,0,0.82}
\definecolor{backcolour}{rgb}{0.95,0.95,0.92}

\lstdefinestyle{mystyle}{
    frame = single,
  commentstyle=\color{codegreen},
  keywordstyle=\color{magenta},
  numberstyle=\tiny\color{codegray},
  stringstyle=\color{codepurple},
  basicstyle=\footnotesize,
  breakatwhitespace=false,         
  breaklines=true,                 
  captionpos=b,                    
  keepspaces=true,                 
  numbers=left,                    
  numbersep=5pt,                  
  showspaces=false,                
  showstringspaces=false,
  showtabs=false,                  
  tabsize=2
}

\begin{document}

% ****************** TITLE ****************************************

\title{BlockLite: A Lightweight Emulator for Public Blockchains}

\numberofauthors{6} %  in this sample file, there are a *total*

\author{
\alignauthor
Xinying Wang\\
       \affaddr{University of Nevada}\\
       \affaddr{Reno, NV, USA}\\
       \email{xinyingw@nevada.unr.edu}
\alignauthor
Abdullah Al-Mamun\\
       \affaddr{University of Nevada}\\
       \affaddr{Reno, NV, USA}\\
       \email{aalmamun@nevada.unr.edu}  
\alignauthor
Hui Lin\\
       \affaddr{University of Nevada}\\
       \affaddr{Reno, NV, USA}\\
       \email{hlin2@unr.edu}  
\and
\alignauthor
Feng Yan\\
       \affaddr{University of Nevada}\\
       \affaddr{Reno, NV, USA}\\
       \email{fyan@unr.edu}
\alignauthor
Mohammad Sadoghi\\
       \affaddr{University of California}\\
       \affaddr{Davis, CA, USA}\\
       \email{msadoghi@ucdavis.edu}
\alignauthor
Dongfang Zhao\\
       \affaddr{University of California}\\
       \affaddr{\& University of Nevada}\\
       \email{donzhao@ucdavis.edu}
}
\date{30 July 1999}

\maketitle

\begin{abstract}
Blockchain is an enabler of many emerging decentralized applications in areas of cryptocurrency, Internet of Things, smart healthcare, among many others.
Although various open-source blockchain frameworks are available,
the infrastructure is complex enough and difficult for many users to modify or test out new research ideas.
To make it worse, many advantages of blockchain systems can be demonstrated only at large scales, e.g., thousands of nodes,
which are not always available to researchers.
This demo paper presents a lightweight single-node emulator of blockchain systems,
namely \mbox{BlockLite}, 
designed to be executing real proof-of-work workload along with peer-to-peer network communications and hash-based immutability.
BlockLite employs a preprocessing approach to avoid the per-node computation overhead at runtime and thus scales to thousands of nodes.
Moreover, BlockLite offers an easy-to-use programming interface allowing for a Lego-like customization to the system, e.g. new ad-hoc consensus protocols.
\end{abstract}

\section{Introduction}

Blockchain, a decentralized and immutable database, has drawn a lot of research interests in various communities,
such as security~\cite{kosba2016-security,jcamenisch_ccs17},
database~\cite{lallen_cidr19,blockbench_sigmod17},
network~\cite{bitcoinng_nsdi16},
distributed systems~\cite{kzhang_icdcs18},
and high-performance computing~\cite{aalmamun_bigdata18}.
Some common challenges shared by these communities include 
(i) the lack of resources to carry out large-scale experiments and 
(ii) much, if not prohibitive, engineering effort to modify sophisticated production (despite open-source) systems to timely test out new ideas.
To this end, multiple blockchain simulators were recently developed,
two of the most popular ones being Bitcoin-Simulator and VIBES.

\textbf{Bitcoin-Simulator}~\cite{agervais_ccs16}
follows the same architecture and protocol of Bitcoin~\cite{bitcoin},
the foremost application in cryptocurrency built upon blockchains. 
Users of Bitcoin-Simulator can specify various protocol and network parameters, 
such as the number of nodes and network bandwidth.
The main goal of Bitcoin-Simulator is to study the trade-off between performance and security.
Because of its design goal, Bitcoin-Simulator simulates the execution of a blockchain network at the block level rather than the transaction level.
Bitcoin-Simulator does not provide a fine-grained control over the application,
limiting its applicability for broader adoption.
In addition, Bitcoin-Simulator simply inserts a series of static time stamps to simulate the proof-of-work (PoW) consensus protocol, 
which does not precisely characterize the behavior of real-world blockchain systems:
for instance, Bitcoin dynamically adjusts the PoW difficulty and the nodes (as known as miners) usually complete the tasks in stochastic time intervals.
Last but not least, Bitcoin-Simulator's network is built upon NS3~\cite{ns3},
a discrete-event network simulator,
which limits the scalability on up to 6,000 nodes. 
Bitcoin network currently consists of more than 10,000 nodes~\cite{bitcoin_scale},
implying that Bitcoin-Simulator cannot simulate the entire network of Bitcoin as of the writing of this paper.

\textbf{VIBES}~\cite{lstoykov_middleware17}
extends Bitcoin-Simulator with the following improvements.
First, VIBES supports a web-based interface for users to visually track the growth of the network.
Second, VIBES improves the scalability of Bitcoin-Simulator by employing a fast-forwarding algorithm,
which essentially designates a coordinator to control the events according to existing nodes' best guess on the block creation time.
Such a centralized coordinator might be acceptable for a single-node simulator at small- or medium-scale,
and yet could be a performance bottleneck for extreme-scale applications.
Similar to Bitcoin-Simulator, VIBES takes the same approach of inserting time stamps to hypothetically carry out the PoW workload.
Both Bitcoin-Simulator and VIBES are coarse estimators of real-world blockchain executions due to the lack of real PoW implementations or a decentralized architecture.

This paper presents \textbf{BlockLite},
the very first blockchain \textit{emulator} designed to be running on a single node with both high scalability and high usability. 
In contrast to Bitcoin-Simulator, BlockLite comprises a module to execute real PoW workload\footnote{Thus making it an \textit{emulator} rather than a simulator},
supports fined-grained transaction management,
and scales out to 20,000 nodes thanks to its efficient network communications built upon distributed queues along with PoW preprocessing that incurs negligible runtime overhead.
Different than VIBES, BlockLite is completely decentralized with no single point of failure or performance bottleneck.
It should be noted that even on a single node the decentralization philosophy of blockchains still holds for a blockchain \textit{emulator} because each user-level thread is now considered as an individual node.
In terms of usability, BlockLite provides an easy-to-use programming interface such that users can programmatically plug in domain-specific components,
such as ad-hoc consensus protocols.

% \don{Add a short paragraph for contributions.}

\figspace
\section{System Design}
\label{sec:design}

The objective of BlockLite is to provide blockchain researchers and practitioners an easy-to-use and lightweight emulator to develop new components and evaluate new ideas,
such as ad-hoc consensus protocols customized for domain-specific applications.
To achieve that objective, BlockLite is designed to be deployed to a single node,
with loosely-coupled components for flexible customization.
In its infrastructure, BlockLite has implemented all the building blocks for a basic blockchain system. 
This section details how these common facilities are designed,
the challenges we encounter, and the approaches we take to build up the emulator.

\begin{figure}[!t]
	\centering
	\includegraphics[width=85mm]{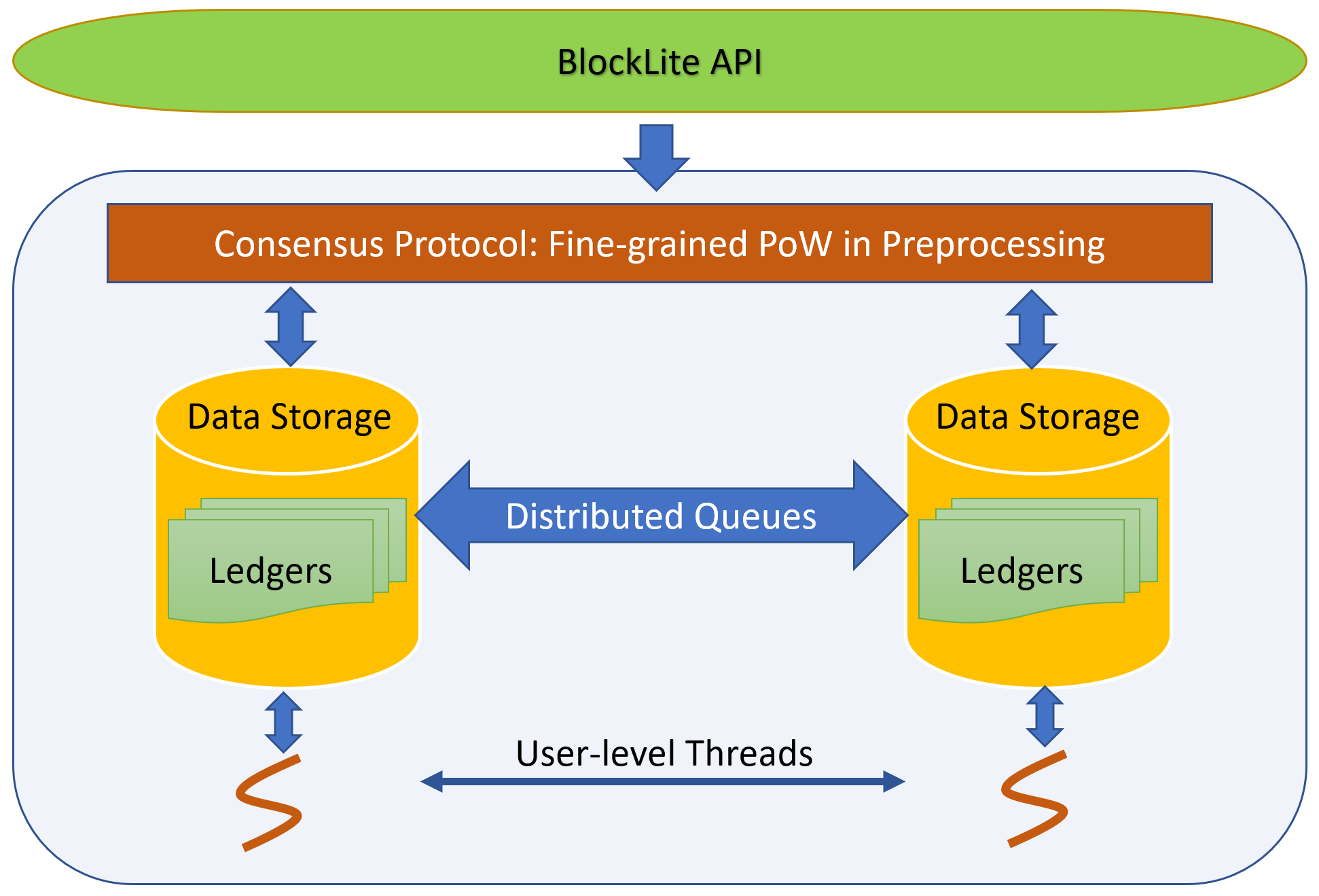}
	\caption{BlockLite Architecture}
	\label{fig:arc}
    \figspace
\end{figure}

The high-level architecture of BlockLite is illustrated in Figure~\ref{fig:arc}.
While the interface, i.e., BlockLite API, will be detailed in~\S\ref{sec:impl},
the infrastructure can be broken down into three categories:
storage, computation, and networking.
In the context of blockchains,
they are usually referred to as \textit{distributed ledgers}, \textit{consensus protocols} (e.g., PoW),
and \textit{network communications}.

\textbf{Distributed Ledgers.} The transaction data of a specific blockchain are replicated, either fully or partially, in distinct files each of which is associated to a hypothetical node. 
The data, also called ledgers, could have been partially duplicated if the following two conditions are satisfied: (i) more than 50\% of nodes have agreed on that the new block (of transactions) is valid and (ii) the current node (other than those nodes who have voted) does not process any request regarding the new block.
    Regardless of specific consensus protocols, a blockchain requires only 50\% votes supporting the new block's validity.
    
\textbf{Consensus Protocols.} A basic proof-of-work protocol is implemented from scratch in BlockLite.
    In contrast to other simulators where the difficulty is \textit{simulated} by time delay and timestamps, 
    BlockLite, as an \textit{emulator}, conducts the real PoW workload by solving the puzzle. 
    The puzzles we define in BlockLite are similar to the Nakamoto protocol in Bitcoin~\cite{bitcoin} in the sense of comparing blocks' hash values against predefined thresholds.
    Nonetheless, BlockLite exhibits an additional feature that preprocesses PoW allowing for fine tuning of puzzle difficulty,
    which is detailed in~\S\ref{subsec:finegrained}.
    
\textbf{Network Communications.} It is one of the most challenging components to emulate the networking in BlockLite that is designed to be working on a single node.
    Fortunately, BlockLite is designed for emulating public blockchains that are based on PoW,
    which is compute-intensive rather than network-intensive\footnote{Private blockchains are indeed network-intensive due to the quadratic number of messages.}.
    Therefore, the real network impact for PoW-based blockchains lies in the network infrastructure's latency rather than bandwidth. 
    BlockLite applies a statistical estimation of time delays for transmitting the messages between nodes,
    each of which is emulated by a user-level thread whose requests are buffered in a distributed queue.
    We will discuss the distributed queue in more detail in~\S\ref{sec:design_network}.
    % It should be clear, however, that this delay should not be confused with other time overhead such as block mining time and block appending time incurred by the application per se.

% With the building blocks aforementioned, BlockLite can serve as a basic blockchain system running in a virtual distributed environment. 
% And yet, there are two additional challenges to be overcome, namely 
% (i) users should be able to efficiently and precisely adjust the puzzle difficulty based on the hardware's computing capability, and
% (ii) users should be able to deploy their new blockchain protocols or applications at a large scale.
% For the first challenge, BlockLite is not based on time stamps but employs a novel method to emulate the puzzle-solving procedure,
% called two-phase zero matching;
% for the second challenge, BlockLite can preprocess the mapping between puzzle's difficulty and the computation time such that large-scale emulations do not involve repetitive computation,
% which in turn achieves a high scalability.

\subsection{PoW Preprocessing for Fine-grained Calibration across Heteregeneous Systems} \label{subsec:finegrained}

% \don{Xinying, we replace this section by describing the new implementation of our PoW.}

One cornerstone of Nakamoto consensus protocol, or any PoW variants, 
is the puzzle-based winner selection:\footnote{As known as ``leader'' in the context of distributed systems}
the hash value of the (block of) transactions is compared against the predefined ``small'' number.
Because BlockLite is designed to be running on an arbitrary node that can be heterogeneous case by case, 
we must provide an efficient yet flexible mechanism to ensure the compatibility across heterogeneous machines. 
To this end, we design BlockLite puzzles as follows.
A puzzle's difficulty is expressed by two sub-fields, $L$ and $M$, in the form of $L.M$ (assuming SHA256~\cite{sha256} is used as the hash function):
$L$ indicates the required number of leading zeros in the 64-hex (i.e., 256-bit) hash value;
$M$ indicates the minimal number of zeros in the middle of the hash value.

\begin{figure}[!t]
	\centering
	\includegraphics[width=80mm]{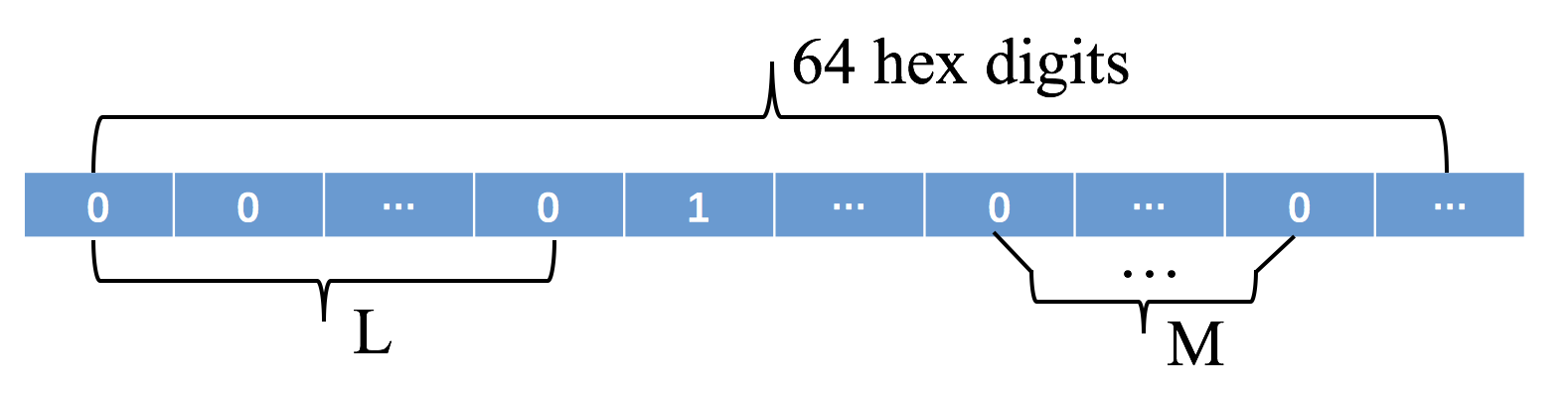}
	\caption{Two-phase Puzzle for Efficient Preprocessing of PoW in BlockLite}
	\label{fig:64hex}
	\figspace
\end{figure}

Figure~\ref{fig:64hex} illustrates how $L.M$ is constructed.
The first part $L$ is semantically equivalent to the Nakamoto protocol:
checking whether a hash value is smaller than a predefined threshold is essentially the same to counting the number of leading zeros in the binary or hex form of the hash value.
$L$ is a coarse-level adjustment of difficulty because the same $L$ might imply a wide spectrum of computation time,
and this is exactly why Bitcoin dynamically adjusts the difficulty every 2016 blocks~\cite{bitcoin}.
To address that, BlockLite introduces the $M$ part to allow the system to check whether there are $M$ zeros in the middle of the hash value satisfying the following conditions:
(i) Any leading zeros in $L$ are not considered;
(ii) Tailing zeros, by definition, are counted towards $M$;
and (iii) Zeros need not be continuous.

The benefit of the additional $M$-zero checking is that we can adjust the puzzle difficulty under the same meta-difficulty, i.e., same $L$ but different $M$'s.
In addition, $M$ is positively correlated to the puzzle difficulty:
a larger $M$ implies more computation time.
To see this, we can think of a larger $M$ representing a super-set of the sets of less zeros with smaller $M$'s.
As a consequence, a smaller $M$ has a higher chance to meet the requirement---the difficulty is lowered.

While the flexibility is significantly improved,
one limitation of this $L.M$ two-phase puzzle is that the two arbitrary difficulty numbers do not follow partial orders in terms of computation time.
That is, if $T(\cdot)$ indicates the computation time of a specific difficulty, it is possible that
\[
T(L_1.M_1) > T(L_2.M_2) \texttt{ and } L_1 < L_2
\]
if $M_1$ is significantly larger than $M_2$.
The root cause of this counter intuition is that $L$ and $M$ are, essentially, incomparable.
For instance, if $L$ is much smaller than $M$,
then finding out $M$ zeros, despite from random positions,
is still much harder than locating a few leading zeros.
As an extreme case, if we have two setups as $L_1.M_1 = 1.63$ and $L_2.M_2 = 2.1$,
obviously the former case is much harder where we will seek for a hash value with all 64 zeros,
as opposed to finding a hash value with two leading zeros and another zero from any of the remaining 62 hex digits.

\subsection{Optimization for Extreme-scale Networking through Distributed Queues}
\label{sec:design_network}

In contrast to existing blockchain simulators, 
BlockLite does not simply insert timestamps for the the completion of PoW;
instead, it solves the real puzzle to accurately \textit{emulate} a real blockchain system.
The downside of this approach is the cost and overhead for large-scale systems.
For instance, Bitcoin has about 10,000 mining nodes as of January 2019~\cite{bitcoin_scale}.
A single machine, despite its multi- or many-cores, is not able to efficiently emulate tens of thousands of nodes each of which works on a compute-intensive puzzle such as finding out a qualified hash value.

BlockLite overcomes the scalability challenge by delegating one node (thread) to solve the puzzle in a preprocessing stage and when the real application runs at a specific difficulty, 
the assigned nodes (or, threads) simply replicate the behavior of the delegation node. 
In doing so, BlockLite achieves the best of both worlds: real execution of PoW and low overhead (i.e., high scalability).
Since the calibration is carried out in a preprocessing state, 
no runtime overhead is introduced. 

The second technique taken by BlockLite to achieve high scalability is the usage of queue-based network communication.
Specifically, we implement a priority queue who manages all the events in the order of their creation time.
That is, the head of the queue always points to the earliest event,
followed by later events each of which is requested by a specific node.
Therefore, the queued events implicitly determine the the orders of nodes completing their tasks (e.g., submitting transactions, solving puzzles, appending blocks),
which significantly reduces the network traffic.

\figspace
\section{Implementation and Interface}
\label{sec:impl}

% \begin{table*}[!th]
% 	\caption{Block Member Variable}
% 	\abdullah{UBlokckID is it a typo?}
% 	\centering
% 	\begin{tabular}{ lll }
% 	\toprule
% 		Variable & Date Type & Meaning	\\
% 	\midrule
% 	uBlokckID & String & ID of the Block \\	\hline
%     creationTime & Timestamp & Creation Time of Current Block \\	\hline
%     creatorID & String  &  Creation ID (Node ID) of Current Block \\	\hline
%     parentBlock &   Block  &  Parent Block of Current Block \\	\hline
%     depth & int & Depth of Current Block \\	\hline
%     previousHash & String &  Hash Value of Parent Block \\	\hline
%     childList &ArrayList<String> & Child List of Current Block \\	\hline
%     numChild & int & Numbers of Current Block Children \\	\hline
%     txnList & ArrayList<Transaction> & Numbers of Current Block Transactions \\	\hline
%     proof & Proovable   & Consensus (PoW by default) \\
%     \bottomrule
% 	\end{tabular}
%     \label{tbl:block}
% \end{table*}

BlockLite is implemented in Java with about 2,000 lines of code.
We have been maintaining a website for the BlockLite project at
\url{https://hpdic.github.io/blocklite};
the source code will be released to \url{https://github.com/hpdic/blocklite}.

Because users' machines are equipped with different resources,
the very first step to deploy BlockLite is to calibrate the parameters in accordance to the system's specification. 
For instance, a throughput of 10 transactions per second might require 7.x difficulty on a high-end server with 32 cores,
and the same throughput might require 4.x difficulty on a mainstream laptop with four cores.
The calibration, also called PoW preprocessing, 
is to allow BlockLite to adjust the difficulty by considering factors input by users (e.g., expected throughput, consensus protocols) as well as system specification (e.g., number of cores, memory size).

When BlockLite runs for the first time, 
it generates a \texttt{difficulty-time} map between difficulty levels and the execution time.
This map is implemented as a HashMap and is accessible to all the nodes.
Whenever a node is waken up according to the consensus protocol,
the node will consult with the difficulty-time map and replay the behavior with controlled randomness.

\begin{minipage}{.95\linewidth}
\begin{lstlisting}[language=Java, caption=BlockLite Plug-in Interface]
public interface Provable { 
    public boolean verifyProof(Block);
    public String generateProof(Block);
}
\end{lstlisting}
\label{lst:interface}
\end{minipage}

\begin{figure*}[!th]
\label{fig:scale}
\centering
% \subfigure[Conventional Blockchain]{
\subfigure[5,000 nodes]{%\raisebox{0mm}{
	\includegraphics[width=55mm]{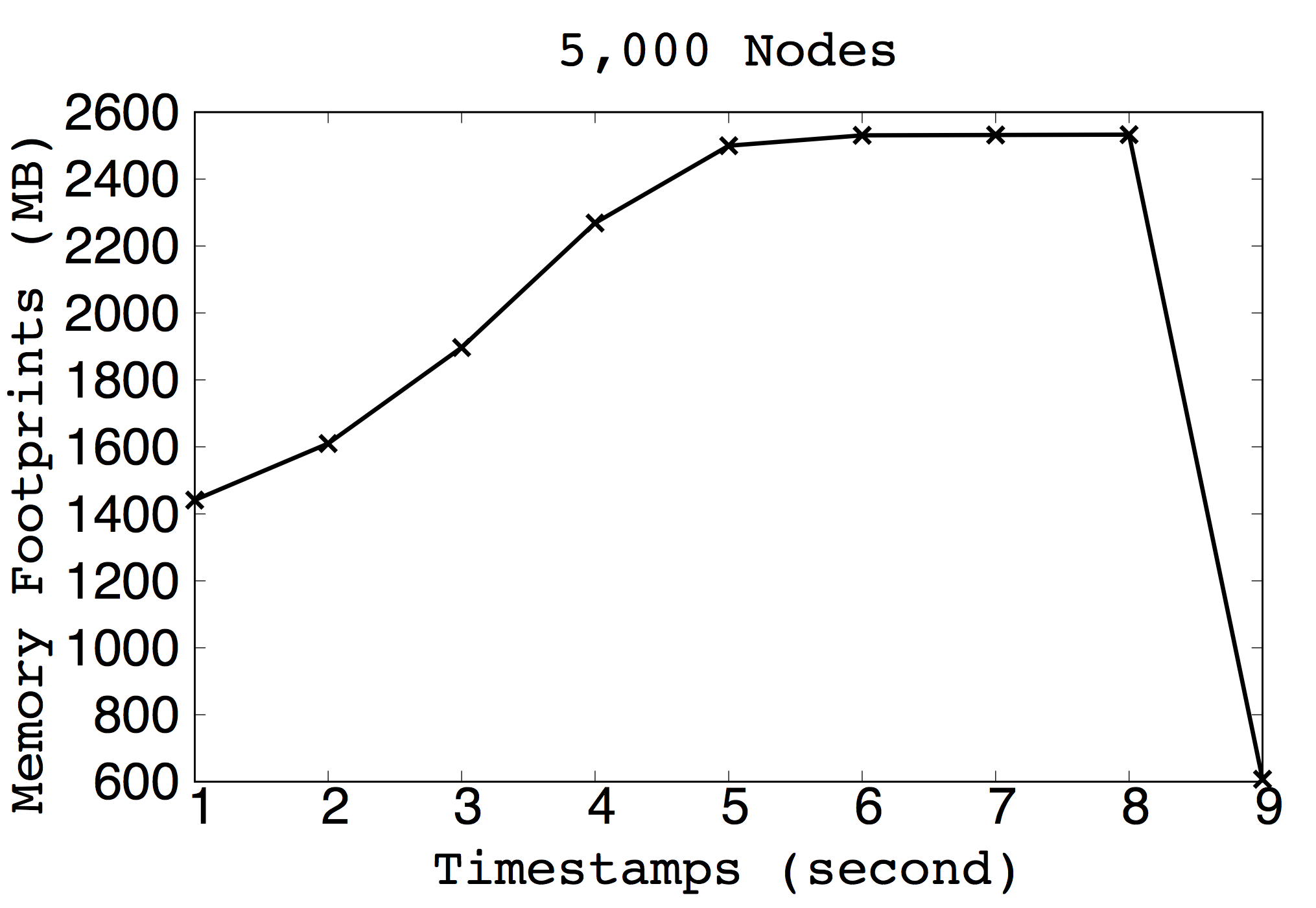}
	\label{fig:5k}
% 	}
}\hspace{3mm}
\subfigure[10,000 nodes]{%\raisebox{8mm}{
	\includegraphics[width=55mm]{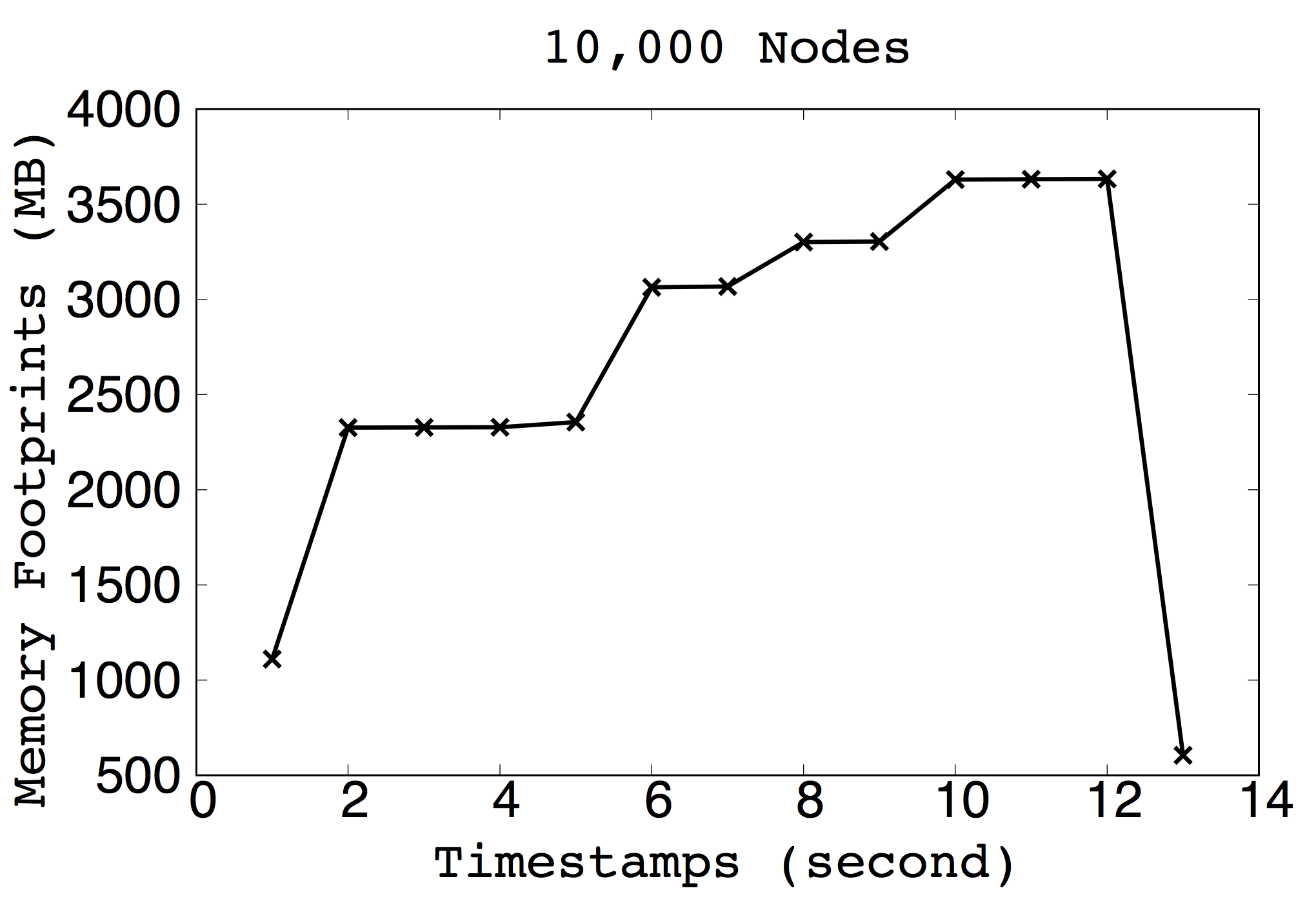}
	\label{fig:10k}
% 	}
}\hspace{3mm}
\subfigure[20,000 nodes]{%\raisebox{8mm}{
	\includegraphics[width=55mm]{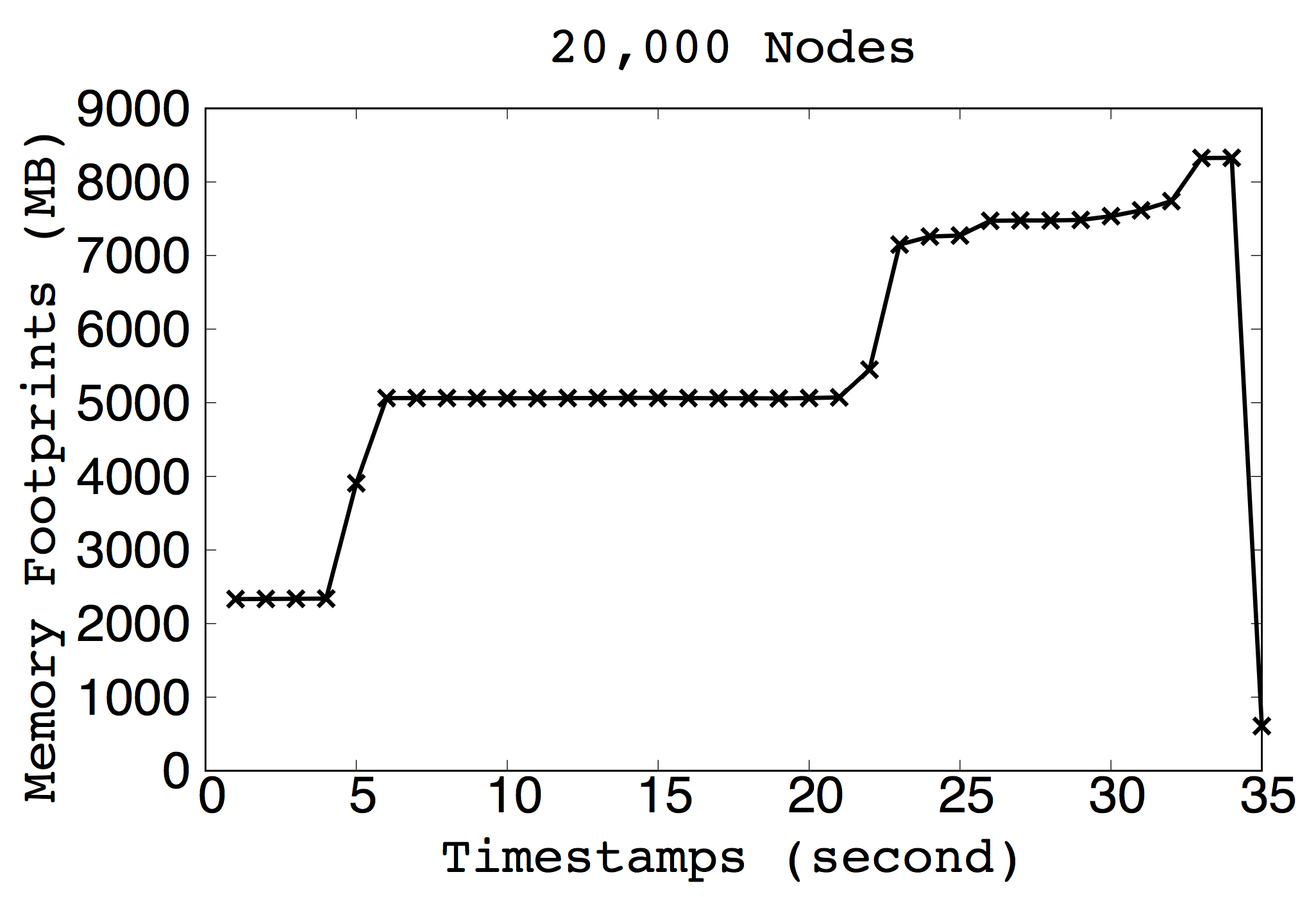}
	\label{fig:20k}
% 	}
}
\caption{Scalability and Memory Footprint of BlockLite}
\figspace
% \label{fig:topo_equality}
\end{figure*}

BlockLite provides an easy-to-use interface for users to plug in application-specific components. 
Listing~\ref{lst:interface} illustrates a simplified code snippet of the interface with two core methods.
Users can implement both methods to inject customized consensus protocols.
For instance, \texttt{generateProof} solves the puzzle;
in PoW, this means to check many nonce numbers until the hash value of the combined data satisfies the condition. 
Both methods take as input a \texttt{Block} object,
whose fields are explained in Table~\ref{tbl:block}.

\figspace
\begin{table}[!h]
	\caption{Block Member Variable}
% 	\abdullah{UBlokckID is it a typo?}
	\centering
	\begin{tabular}{ lll }
	\toprule
		Variable & Meaning	\\
	\midrule
	blockID & ID of the Block \\	\hline
    creationTime & Creation Time of Current Block \\	\hline
    creatorID & Creation ID (Node ID) of Current Block \\	\hline
    parentBlock &  Parent Block of Current Block \\	\hline
    depth & Depth of Current Block \\	\hline
    previousHash &  Hash Value of Parent Block \\	\hline
    childList & Child List of Current Block \\	\hline
    numChild & Numbers of Current Block Children \\	\hline
    txnList & Numbers of Current Block Transactions \\	\hline
    proof & Consensus protocol (Nakamoto by default) \\
    \bottomrule
	\end{tabular}
	\figspace
    \label{tbl:block}
\end{table}

\section{Demo Scenario}

We will demonstrate BlockLite on Amazon EC2.
The BlockLite VM image will be made public such that audience can launch their own BlockLite instances when attending our demo presentation.
Due to the time limit and cost, we will launch a \texttt{t2.2xlarge} EC2 instance and run the emulation on hundreds of nodes.
We have tested BlockLite on up to 20,000 nodes on a test bed with 48 AMD EPYC cores and 192 GB memory.
We report the results on this test bed running over one million transactions in the following.

Figure~\ref{fig:scale} shows BlockLite's real-time executions, along with memory footprint, on 5,000, 10,000, and 20,000 nodes, respectively. 
The puzzle difficulty is set to one for the sake of fast demonstrations.
We can observe that even at the the real scale of Bitcoin---10,000 nodes,
BlockLite can finish the emulation in 13 seconds with reasonable memory footprint of less than 4 GB.

\begin{figure}[!t]
	\centering
	\includegraphics[width=86mm]{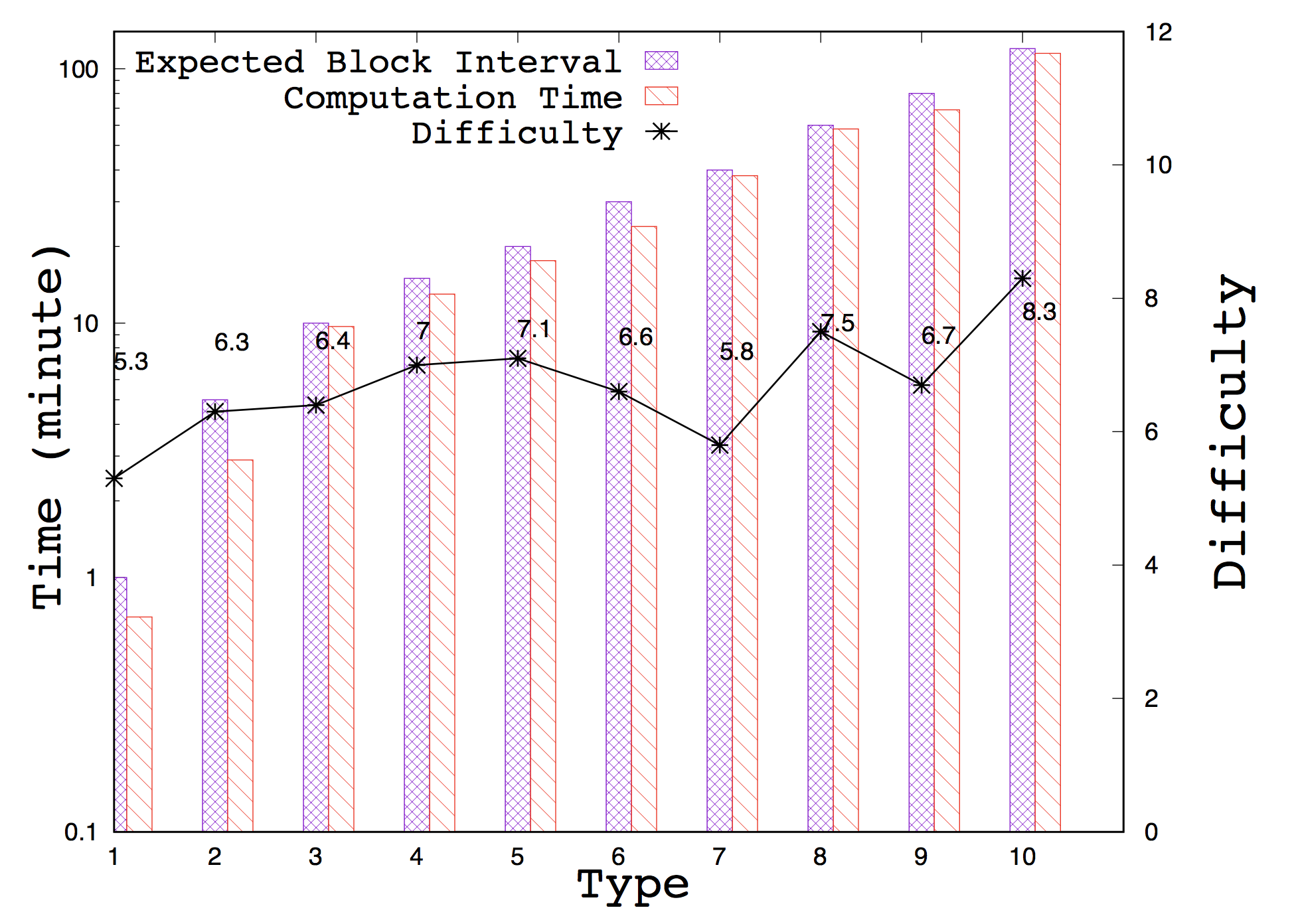}
	\caption{Block Creation Rate with Various Calibration Difficulties}
	\label{fig:diff}
    \figspace
\end{figure}

Figure~\ref{fig:diff} shows the calibration map between the fine-grained difficulty and the emulated block creation rate.
For practical time intervals,
e.g., ten minutes or more, 
the emulated processing times are close to the expected time intervals.
We also plot the preprocessing difficulties in the figure;
the result suggests that most difficulty values range in between five and eight,
covering mining time from 1 to 100 minutes.

% Figure~\ref{fig:scalability} shows the emulation time on various scales,
% with difficulty set to five.
% BlockLite is expected to run for a longer time with more participating nodes as the limited resources serve more hypothetical nodes.
% The amortized response time,
% as plotted on the right Y-axis,
% is in the order of a few seconds,
% which are completely acceptable for emulating purposes.

% \begin{figure}[!t]
% 	\centering
% 	\includegraphics[width=80mm]{scalability.png}
% 	\caption{Emulation Time at Various Scales}
% 	\label{fig:scalability}
%     \figspace
%     \vspace{-1mm}
% \end{figure}

% \begin{figure}[!t]
% 	\centering
% 	\includegraphics[width=80mm]{5k.png}
% 	\caption{Memory Footprints of 5 thousands nodes}
% 	\label{fig:5k}
%     \figspace
% \end{figure}

% \begin{figure}[!t]
% 	\centering
% 	\includegraphics[width=80mm]{10k.png}
% 	\caption{Memory Footprints of 10 thousands nodes}
% 	\label{fig:10k}
%     \figspace
% \end{figure}

% \begin{figure}[!t]
% 	\centering
% 	\includegraphics[width=80mm]{20k.png}
% 	\caption{Memory Footprints of 20 thousands nodes}
% 	\label{fig:20k}
%     \figspace
% \end{figure}

\figspace
\section{Conclusion}

This paper introduces BlockLite, the very first decentralized emulator for public blockchains scalable to 20,000 nodes.
BlockLite achieves such a high scalability through an offline calibration of PoW execution and distributed queues. 
In terms of usability, BlockLite provides an easy-to-use interface to plug in application-specific components such as ad-hoc consensus protocols.
The demo will showcase BlockLite's programming interface,
its offline calibration of preprocessing PoW, 
and the emulation of one million transactions at large scales.
% We will also briefly introduce how BlockLite is being used as a major evaluation tool for various ongoing Blockchain-related projects.

\figspace
\section*{Acknowledgement}
% \noindent
% \textbf{Acknowledgement.} 
This work is in part supported by NASA and Google.

{\footnotesize
\bibliographystyle{IEEEtran}
\bibliography{ref_new}
}

\end{document}